\documentclass[a4paper,12pt]{article}
\usepackage{amsfonts}
\usepackage{amssymb}
\usepackage{amsmath}
\usepackage{graphicx}
\usepackage{placeins} 
\usepackage{lscape}
\usepackage[round,authoryear]{natbib}
\usepackage{color}
\usepackage{eurosym}
\usepackage{longtable}

\renewcommand{\euro}{}
\newcommand{\dsum}{\sum}
\newcommand{\sym}[1]{\rlap{#1}}
\newtheorem{definition}{Definition}

\begin{document}

\title{Key drivers of EU budget allocation: Does power matter?}
\author{Vera Zaporozhets\thanks{%
Corresponding Author. Toulouse School of Economics (LERNA, INRA), e-mail:
vzaporoz@toulouse.inra.fr} \and Mar\'{\i}a Garc\'{\i}a-Vali\~{n}as\thanks{%
University of Oviedo} \and Sascha Kurz\thanks{%
University of Bayreuth}}
\date{}
\maketitle

\begin{abstract}
We examine the determinants of the EU budget expenditures allocation among
different countries. In line with earlier literature, we consider two
alternative explanations for the EU budget distribution: political power vs.
\textquotedblleft needs view.\textquotedblright\ Extending the original data
set from \cite{kauppiwidgren04}, we analyze the robustness of their
predictions when applying a different measure of power and more
sophisticated econometric techniques. We conclude that the nucleolus is a
good alternative to the Shapley-Shubik index in distributive situations such
as the case of EU budget allocation. Our results also show that when
explaining budget shares, the relative weight of political power based on
the nucleolus is lower than the predictions of previous studies based on the
Shapley-Shubik index.


\end{abstract}

\pagebreak

\section{Introduction}

In 2013, the European Union (EU) expenditure budget was around  \euro{} 149
billion, with cohesion, and agricultural and environmental resources being
the primary EU policies, with shares of 46.8\% and 39.8\% respectively. Due
to the magnitude of these figures, the distribution of the EU budget among
different countries is a crucial issue to analyze. In particular, we focus
on the relative weights of different factors when explaining the budget
shares corresponding to each EU member.

Previous literature 
\citep{courcheneetal93, andersontyers95, tangermann97,
kandogan00, kauppiwidgren04, kauppiwidgren07} has tested two alternative
explanations of the EU budget distribution across the members states. The
first is a \textquotedblleft needs view,\textquotedblright\ which states
that the budget allocation is determined by the principles of solidarity.
According to this hypothesis, the countries with a high agriculture sector
weighting and/or a relatively worse economic situation emerge as the major
recipients of the EU budget. In fact, some of the previous studies have
focused exclusively on this explanation 
\citep{courcheneetal93,
andersontyers95, tangermann97}. The second explanation is that budget
allocation across the members is reflected by the distribution of their
political power. Thus, those countries with more power in the allocation
process could receive larger shares of the budget.

Some studies combine both: needs and the power views 
\citep{kandogan00,
kauppiwidgren04, kauppiwidgren07, aksoy10}. Thus, empirical analysis by \cite%
{kauppiwidgren04} shows the strong prevalence of political power motives.

Their results indicate that political power has much higher weight than
needs when determining the allocation of budget expenditures among EU member
states.

The overall purpose of this paper is to reconsider the analysis of \cite%
{kauppiwidgren04} and to challenge their conclusions. To do so, we extend
the original data set (1976-2001) up to 2012 and introduce alternative model
specifications. In contrast to the existing studies which have used the
Shapley-Shubik index as a measure of political power, we employ a different
measure; the nucleolus. It has been argued that the nucleolus is an
appropriate power measure in distributive situations as well as a good
alternative to traditional measures, such as the Shapley-Shubik index.%
\footnote{%
For example, \cite{montero05}, \cite{montero13} and \cite{lebreton12b} among
others.} Moreover, we apply sophisticated econometric techniques, which are
more suitable for the analysis.

The paper is structured as follows. Section 2 provides a brief introduction
on the EU budget, including the processes of designing and allocating EU
expenditures and revenues. In Section 3, we discuss the theoretical
properties of two different power indices. Specifically, we highlight the
advantages of the nucleolus over other indices in distributive situations,
such as the EU budget allocation. Finally, in Section 4, we specify a simple
empirical model in order to determine the key drivers for EU budget
allocation. Section 5 concludes with a summary of the main findings in
addition to some policy implications.

\section{EU budget: procedure and evolution}

As mentioned in the Introduction, the EU expenditure budget represents a
significant amount of resources. In 2013, total expenditures were \euro
148,468 million. Although this is not a substantial amount in relative terms
(just 1.13\% of the EU-27 Gross National Income, GNI), some crucial policies
were developed using EU funding. Examples are the Common Agricultural Policy
(now part of a more extensive section on the preservation and management of
natural resources) or several policies oriented towards the economic
development of some target regions (cohesion and competitiveness policies).
Each EU member also has to contribute to the EU budget, by means of
GNI-based resources (74.3\%), VAT-based own resources (9.5\%), and
traditional own resources (TOR, 10.4\%).

In 1976, the EU expenditure budget amounted to \euro 7,563 million. In the
last decades, the EU budget has been growing.\footnote{%
Figures on the EU budget are available at
http://ec.europa.eu/budget/figures/index\_n.cfm.} This increasing path can
be interpreted as a snap that simultaneously captures the history of EU
integration and several budgetary reforms. Regarding the enlargement
process, there are some significant facts which could have an impact on the
evolution of the EU expenditures. In 1986, the EU grew from 10 to 12
countries, through the integration of Portugal and Spain as new members.
Similarly, Austria, Finland and Sweden joined the EU/EC in 1995.
Furthermore, one of the largest phases of expansion occurred in 2004, when
the EU grew from 15 to 25 member countries.\footnote{%
The new members were the Czech Republic, Cyprus, Slovak Republic, Slovenia,
Estonia, Hungary, Latvia, Lithuania, Malta and Poland.}

Successive EU reforms have changed the structure of the budget. In this
respect there are some facts that are worth mentioning. At the Brussels
European Council in February 1988, a political agreement on doubling, in
real terms, the budget of the Structural Funds between 1987 and 1993 was
reached. Subsequently, Member States agreed at the Edinburgh European
Council in December 1992 that the budget for structural operations would be
further increased, specifically for the cohesion countries (Greece, Ireland,
Portugal and Spain). Also in Edinburgh, Member States decided to strengthen
some particular policies, such as research and development, external actions
and financial aid to Central and Eastern European countries. Although there
were several agreements on setting budgetary limits to the growth rate of
expenditure, the basis of a stringent budgetary discipline was established
in the Agenda 2000 agreements. These reforms have all had an impact on the
level and the structure of budget expenditures, and have led to some changes
in the accounting system. The budget has therefore undergone some structural
reforms, the most significant being those of 1992 and 2006.

Regarding the procedure for elaboration and approval of the EU budget, there
are several institutions involved. The European Commission, the Council and
the Parliament participate in the process of elaborating the EU budget.
However, over the past decades, the role of each institution, as well as the
voting rules, have undergone a number of changes \citep{kauppiwidgren07}.
The relationship between the Council and the Parliament was controversial
until the 1992 Edinburgh meeting, where an Interinstitutional Agreement
between both institutions was established in order to facilitate the process
of making budgetary decisions.

The budget elaboration process involves the following steps. First, based on
the multi-annual financial framework in force and the budgetary guidelines
for the coming year, the European Commission prepares a preliminary draft
budget. Within this stage, spending priorities are established, plus caps or
ceilings to limit the maximum growth rate of different budgetary sections
and the total budget.

Once a preliminary draft is drawn up, the European Commission submits it to
the Council and the Parliament. The budgetary authority, comprising both
institutions, amends and subsequently adopts the draft budget. The Council
is then expected to adopt its position on the preliminary draft budget
proposed by the Commission, elaborate and approve a definitive draft budget.
Next, the Parliament can modify the draft, by adopting amendments to the
Council's position, or by proposing some amendments to particular expenses.
The final proposed document is then approved by simple majority by the
Parliament. Following this, the Council has a second reading of the
document, adopting it by a larger majority than that required at the
Parliament.\footnote{%
Usually, at least a qualified majority is required to adopt budgetary
decisions at the Council.} A second reading by the Parliament and the
definitive adoption marks the end of the process.

In this paper, we consider the voting decisions of the Council %
\citep{bindseilhantke97, kauppiwidgren04}. Although the Parliament has
recently increased their weight in the EU decision process,\footnote{%
The Treaty of Lisbon extended the role of the Parliament. It was signed by
the EU member states on 13 December 2007 and entered into force on 1
December 2009. From that moment, European Parliament could decide on both
compulsory and non-compulsory expenses, extending its power and
responsibilities with regard to the budget making process. With the entry
into force of the Treaty of Lisbon a new system known as a \textquotedblleft
double majority\textquotedblright\ was introduced. It entered into force on
1 November 2014. The Nice system remained applicable during the transition
period up to 31 October 2014. For an interesting discussion on the Treaty of
Nice, see \cite{heinemann03}.} some EU institutional features have supported
the approach based on the voting framework at the Council.\footnote{%
Among others, the approval procedure applied during the period analyzed in
this research (1976-2012) and the qualified majority required at the Council
to approve the final EU budget.}

\section{Power indices: the nucleolus versus the Shapley-Shubik index}

In recent decades, there has been a growing literature, both theoretical and
applied, on power measures. However, as yet, there is no consensus as to the
best way to measure power. While analyzing the distribution of the EU budget
among different countries, previous studies have applied the Shapley-Shubik
index (SSI) \citep{kauppiwidgren04}, one of the mostly commonly used power
measures in this context. By contrast, in this study we propose an
alternative measure, the nucleolus\footnote{%
Thus, \cite{felsenthal98} argue that the SSI is a measure of
\textquotedblleft P-power,\textquotedblright\ where P stands for
\textquotedblleft purse,\textquotedblright\ and it evaluates a voter's
expected relative share of a fixed budget. As we argue in the paper (see
also \cite{lebreton12b}) the nucleolus can also be considered a power
measure in the distributive setting, and can be a good alternative to the
SSI. Consequently, in this paper we focus on these two measures. At the same
time, the SSI can also be considered a measure of \textquotedblleft
I-power\textquotedblright\ \citep{felsenthal98}, which assesses the
probability of a player casting a decisive vote. Other measures of I-power
include the Banzhaf index, the Johnston index, and the Deegan-Packel index.} %
\citep{schmeidler69}. In the following, we provide strong arguments to
support our choice. In the subsequent section, we compare how the two
indices perform in practice, and analyze whether the conclusions reached by 
\cite{kauppiwidgren04} are robust with regard to their choice of power index.

The general discussion on which power measure is best, and which properties
it should possess, remains open. \cite{napel04} therefore divide existing
studies on power indices into two \textquotedblleft disjoint methodological
camps,\textquotedblright\ and propose a unified framework to link them:
\textquotedblleft On the one hand, such a framework should allow for
predictions and ex post analysis of decisions based on knowledge of
procedures and preferences. On the other hand, it must be open to ex ante
and even completely a priori analysis of power when detailed information may
either not be available or should be ignored for normative
reasons.\textquotedblright\ We address the discussion in a specific
distributional setting, that is, the allocation of a fixed budget across the
members of an organization, with the key preference assumption being that
each member cares only about their own share.

Following \cite{napel04}, let us consider two requirements in turn. First,
the power measure should be based on the explicit decision-making procedures
and the knowledge of the preferences. To this end, it is important that the
political analysis takes into account game forms. In this respect, both the
SSI and the nucleolus are suitable measures to analyze bargaining situations
such as the distribution of the EU budget. Each of the two measures are
founded on a non-cooperative framework, in that either of them arises as a
payoff from a well-specified bargaining game.\footnote{%
For instance, \cite{gul89} constructs a non-cooperative game mimicking
bargaining process in the markets. One of the main results state that the
payoffs associated with efficient equilibria converge to the agents' Shapley
values as the time between periods of the dynamic game goes to zero. Even
though Gul's bargaining procedure is very natural, his results are not
relevant for majority games. Some examples of a less natural bargaining
procedure but more general results are \cite{hartmascolell96} and \cite%
{vidalpuga08}. As for the nucleolus, it has been proved to correspond to the
vector of expected payoffs in the legislative bargaining game with random
proposers according to \cite{baron89}, where voters directly put forward
proposals and vote over the division of a budget. If proposal probabilities
coincide with the nucleolus, then the nucleolus is the unique vector of
expected payoffs \citep{montero06}. The equality of the expected payoffs to
the nucleolus also holds for other proposal probabilities depending on the
voting game.}

According to the second requirement, one would not want the power analysis
to be extremely sensitive to the details of the game form used to describe
the non-cooperative decision process. In the following, we show that only
the nucleolus passes this test.

In order to encompass the idea of a robust power measure in our specific
distributional framework, we address the bargaining set, a solution concept
for coalitional games \citep{maschler13}. The idea behind the bargaining set
is that when the players decide how to divide the worth of the coalition,
the player who is not satisfied with the proposed share may object to it.
The objection goes against another player, calling for this player to share
their part with the objecting one. The player against whom the objection is
made may (or may not) have a counter objection. An objection which does not
have a counter objection is called justified. The bargaining set consists of
all imputations in which no player has a justified objection against any
other player.

It seems that the bargaining set properly describes the decision-making
procedure within EU institutions (see Section 2). Additionally, one of the
properties of the bargaining set is that, contrary to the core, it is never
empty. However, the bargaining set is often large, in which case there is
the problem of choosing a unique outcome. In such cases, the nucleolus is a
good candidate, since it always exists, it is unique and it belongs to the
bargaining set. On the contrary, in general the Shapley value is not in the
bargaining set. The following example supports this argument. Let us
consider three individuals with individual one being a vetoer. This means
that a decision is passed only when player one is present in a group voting
for the decision, however if they are on their own, they cannot get the
decision passed. In such a situation, the core, the nucleolus, and the
bargaining set coincide, and attribute the whole surplus to player one. On
the contrary, the Shapley-Shubik index is $(2/3,1/6,1/6)$. One may check
that under the distribution according to the Shapley-Shubik index, player
one has an objection. For example, player one can offer player two to share
the part of player three. In this setting, player one has a lot of power,
and only one extra vote is needed to validate the decision. The nucleolus
models the process of Bertrand competition between players two and three.

To summarize, both the SSI and the nucleolus have foundations in
non-cooperative bargaining games, which make them suitable for ex post
political analysis. However, only the nucleolus satisfies the requirement of
being a measure that is open to an ex ante analysis of the distributive
situations, as stated in \cite{napel04}. Given our specific framework, these
arguments allow us to favour the nucleolus versus the SSI in the empirical
analysis of the EU budget distribution. In the Appendix, we provide the
formal definitions for the SSI and the nucleolus, as well as the figures for
both power measures for the period 1958-2012.

\subsection{Example}

In this subsection we provide computations of the SSI and the nucleolus for
the first EU Council of Ministers (1958 - 1972). During that period the
Council consisted of representatives from six countries. The three
\textquotedblleft big\textquotedblright\ countries (Germany, Italy and
France) held four votes each, the two \textquotedblleft
medium\textquotedblright\ countries (Belgium and the Netherlands) held two
votes each and the \textquotedblleft little\textquotedblright\ country
(Luxembourg) held one vote. A qualified majority was set at $12$ out of $17$%
, i.e., passing a decision required at least $12$ votes in favour of the
decision. As has been highlighted in a number of studies,\footnote{%
For example, \cite{felsenthal97}, among others.} Luxembourg was powerless in
such a situation. Since other member states held an even number of votes,
Luxembourg was never formally able to make any difference in the voting
process. The results are summarized in Table 1.

\begin{table}[h] \centering%
\caption{\bf The Council of Ministers (1958 - 1972).\label{key}}\label%
{Table1}%
\begin{equation*}
\begin{tabular}{lrrr}
\hline\hline
\textbf{Country} & \textbf{Weight} & \textbf{SSI } & \textbf{Nucl} \\ \hline
Germany & \multicolumn{1}{l}{$4$} & $0.233$ & $0.250$ \\ 
Italy & \multicolumn{1}{l}{$4$} & $0.233$ & $0.250$ \\ 
France & \multicolumn{1}{l}{$4$} & $0.233$ & $0.250$ \\ 
Belgium & \multicolumn{1}{l}{$2$} & $0.150$ & $0.125$ \\ 
Netherlands & \multicolumn{1}{l}{$2$} & $0.150$ & $0.125$ \\ 
Luxembourg & \multicolumn{1}{l}{$1$} & $0$ & $0$ \\ \hline
\textit{Quota} & \multicolumn{1}{l}{$12$} &  &  \\ 
Total votes & \multicolumn{1}{l}{$17$} &  &  \\ 
\textit{Quota (\%)} & \multicolumn{1}{l}{$70.59$} &  &  \\ \hline
\end{tabular}%
\end{equation*}%
\end{table}%

According to the nucleolus, a \textquotedblleft medium" country receives
half as much weight as a \textquotedblleft big\textquotedblright\ country.
This is quite intuitive, since in a minimal winning coalition,\footnote{%
A minimal winning coalition is a group of countries whereby if they all vote
in favour of a decision, it is passed. Furthermore, none of the countries
can be excluded, i.e., if one of the countries change the vote from
\textquotedblleft yes\textquotedblright\ to \textquotedblleft
no\textquotedblright \thinspace\ the decision can no longer be passed. In
this scenario, there are two types of the minimal winning coalitions: three
"big" countries, or two \textquotedblleft big\textquotedblright\ countries
and two \textquotedblleft medium\textquotedblright\ countries.} a
\textquotedblleft big\textquotedblright\ country can be replaced by two
\textquotedblleft medium\textquotedblright\ ones. Such substitutability
often holds for the nucleolus in contrast to other power indices, but it
does not hold in all cases.\footnote{%
For a more detailed discussion see, for example, \cite{montero05}.} As a
consequence, in this case the nucleolus treats all minimal winning
coalitions equally. It prescribes the total \textquotedblleft
wealth\textquotedblright\ for both types of coalitions as being equal to $%
0.75$. In contrast, according to the SSI, the minimal winning coalitions of
the first and the second type get different values, $0.766$ and $0.7$
respectively.

We now highlight another interesting feature of the nucleolus. In 1973, when
compared to 1958, the \textquotedblleft big\textquotedblright\ countries get
the same power according to the nucleolus. However, other countries, even
though they are not dummies, get zero. This is impossible for the SSI or
other power indices, but it is not unusual for the nucleolus.\footnote{%
See, for example, \cite{montero05}.} As a result, the nucleolus is very
different from the SSI and other indices in this example.

\section{Empirical Application}

\subsection{Data and empirical model}

As explained in the Introduction, this paper aims to identify significant
key drivers and trends for EU budget allocation. In order to discuss the
findings of \cite{kauppiwidgren04}, we extend their data set to include
observations for the period 1976-2012. Interestingly, this period covers
different phases of EU integration: from 1976 to 1980 (EU9), from 1981 to
1985 (EU10), from 1986 to 1994 (EU12), from 1995 to 2003 (EU15), from 2004
to 2006 (EU25) and from 2007 to 2012 (EU27).\footnote{%
Although data for 2013 are available, this last year was removed from the
dataset. This is the first year with EU-28 with Croatia as a new EU member,
so in the context of unbalanced panel data methods, this observation would
need to be dropped since there is only one period for that observation %
\citep{bluhm2013, Wooldridge2010b}.}

In this respect, a general model will be proposed, where the budget share on
the whole EU budget of each country depends on an index of political power
and a set of variables representative of budgetary needs. The empirical
model is presented as follows:

\begin{equation}
b_{it}=f(p_{it},Z_{it})+u_{it},
\end{equation}

where $b_{it}$ is the percentage of the total EU expenditure budget
allocated to country $i$ in the year $t$, $p_{it}$ is an index of political
power for country $i$ and period $t$, and $Z_{it}$ is a vector of factors
representative of country $i$'s needs in period $t$. Finally, $u_{it}$
represents the error term.

We have proposed two alternative specifications of the dependent variable,
following the procedure suggested by \cite{kauppiwidgren04}. On the one
hand, we consider the total expenditure budget share that each country gets
in the negotiation process ($exp$). On the other hand, an alternative
variable is defined, introducing an adjustment to take into account the UK's
budget rebate and other similar compensations ($exp_{adj}$).\footnote{%
This rebate was a compensatory payment made to the UK government in 1985.
The main argument in the rebate negotiations was that a high proportion of
the EU budget was spent on the Common Agricultural Policy (or CAP), which
benefits the UK much less than other countries, as it has a relatively small
farming sector as a percentage of GDP. The compensation consists of
reallocating some of the original UK monetary contributions to be paid by
the remaining member states. Additionally, some minor compensation payments
received by other member countries (e.g. Sweden and the Netherlands) are
also included in the calculations.}

We also use some of the original variables proposed by \cite{kauppiwidgren04}
as independent variables. First, two different alternatives to measure
political power discussed in the previous section are included in the
analysis;\footnote{%
Several alternative political power indices have been considered in the
estimations, such as Banzhaf, Johnston, Public Good, and Deegan-Packel
indices, see \cite{kurz2014} for a recent overview on power indices.
However, none of these power indices improved the explanatory power shown by
the Shapley-Shubik or the nucleolus. The Banzhaf and the Johnston indices
show similar levels of the adjusted $R^{2}$ while there are plenty of
independent variables that are not significant. The Public Good and the
Deegan-Packel indices seem to be more sensitive to changes in the model
specification.} namely, the SSI ($p_{ssi}$) and the nucleolus ($p_{nucl}$).
The latter power index was not originally included by \cite{kauppiwidgren04}%
, but has been considered for a comparison to be made with the SSI.
Additionally, needs are shown using a set of variables ($Z$): each country's
share of the total agricultural production ($agri$), and the ratio of each
country's GDP per capita and the EU wide GDP per capita ($income$). Table 2
shows some descriptive statistics of the main variables:\footnote{%
Budget shares have been calculated using the information taken from the
European Commission financial reports. The remaining data were taken from
the Eurostat statistics website. Political power indices have been
calculated 
as described in Appendix 1.}

\bigskip \FloatBarrier
\begin{table}[h]
\caption{Summary statistics }
\label{sumstat}\centering
{\normalsize {\small \centering
\begin{tabular}{lcccc}
\hline\hline
\multicolumn{1}{c}{\textbf{Variable}} & \textbf{Mean} & \textbf{Std. Dev.} & 
\textbf{Min.} & \textbf{Max.} \\ \hline
exp & 0.0583 & 0.0530 & 0.0002 & 0.2256 \\ 
$\text{exp}_{\text{adj}}$ & 0.0584 & 0.0529 & 0.000 & 0.2256 \\ 
$\text{p}_{\text{ssi}}$ & 0.0643 & 0.0474 & 0.0081 & 0.1786 \\ 
$\text{p}_{\text{nucl}}$ & 0.0643 & 0.0641 & 0.000 & 0.2500 \\ 
agri & 0.0643 & 0.0720 & 0.0004 & 0.3383 \\ 
income & 1.0000 & 0.3214 & 0.4087 & 2.6786 \\ \hline
\end{tabular}
} }
\end{table}
\FloatBarrier

One may observe that the SSI shows higher dispersion levels as compared to
the nucleolus. The average expenditure budget percentage received is around
6\%. It is also worth mentioning that the variables representing budget
needs present high levels of dispersion. Thus, country members are
heterogeneous in terms of their economic structure.

\bigskip

\subsection{Results}

In order to carry out a sensitivity analysis, we have proposed four
different specifications. Estimates appear in Tables 3-7. The four
specifications are the result of combining two different dependent variables
($exp$ in Tables 3 and 4; $exp_{adj}$ in Tables 5 and 6) with the two
political power indices described earlier ($p_{ssi}$ in Tables 3 and 5; $%
p_{nucl}$ in Tables 4 and 6). Finally, Table 7 provides the basis for a
comparison of the four specifications, as marginal effects are reported.

Regarding the econometric techniques, we have considered several models.
First, we keep the pooled baseline Ordinary Least Squares specification
(OLS) originally proposed by \cite{kauppiwidgren04}, in order for it to be
compared with more sophisticated techniques. The analysis presented in the
current paper suggests unobservable heterogeneity due to the strong
differences among country members from different perspectives. Moreover,
since the dependent variable is a share, econometric methods should be
adapted to take this into account. Thus, OLS seems to be a non-robust
econometric technique in this context.

In order to resolve any issues associated with the OLS method, two
fractional methodologies have been proposed. First, a Generalized Linear
Model (GLM) based on a probit distribution has been applied. Second, since
our panel data is clearly unbalanced, an alternative fractional model based
on probit distribution has been considered (FHETPROB). Note that the nature
of unbalancedness could require models that explicitly allow for
heteroskedasticity \citep{Wooldridge2010b,
Wooldridge2010a}. In both cases, a clustered option has been used to
estimate the variance$-$covariance matrix.

Additionally, Equation (1) has been extended to include a set of variables
that consider the effect of the EU enlargement ($EU$). In this respect, a
set of dummy variables has been defined: $EU10$, $EU12$, $EU15$, $EU25$, $%
EU27$. Those variables take the value of 1 when the number of country
members is 10, 12, 14, 25 and 27 respectively, and 0 otherwise. Furthermore,
some interactions of these dummy variables with the political power indices
have been considered, since there could be different impacts of power
depending on the number of countries integrated into the EU. The dummy
variables have been included in OLS (denoted by $OLS_{d}$), GLM and FHETPROB
models.

Moreover, additional dummy variables have been generated %
\citep{Wooldridge2010a} in order to capture the panel unbalancedness
structure in the context of the FHETPROB specification. Assuming that the
global panel data set is composed of different subpanels $T_{i}$, both the
outcome and variance equation are allowed to depend on the number of
observations in each subpanel. The new dummy variables are denoted by $%
tobs32 $, $tobs27$, $tob18$, $tobs9$, $tobs6$, and take the value 1 when the
country is observed for 32, 27, 18, 9 and 6 years respectively, and 0
otherwise.

The results show some general facts that are observed in the majority of
cases. Both power and needs are significant key drivers of budget
allocation. Thus, the higher the political power, the higher the expenditure
share. Additionally, those countries with more intensive agricultural
activity and lower relative income emerge as the beneficiaries of EU
policies, as they receive higher shares of the overall budget.

Regarding the econometric models, and comparing both fractional techniques,
it seems that FHETPROB enables an encreased significance of the three main
variables related to power and needs. The inclusion of a variance equation
based on the unbalancedness structure helps to refine the results. In all
the specifications, the majority of variables explaining the variance are
highly significant.

When comparing the performance of different models and when focusing on
alternative power indices, there are also interesting findings. In terms of
OLS regressions, the adjusted $R^{2}$ values show that models based on $%
p_{ssi}$ have a higher explanatory power than those based on $p_{nucl}$.
However, the differences are considerably smaller when dummy variables
representative of EU enlargement process are included in the analysis. Thus,
both power indices seem to perform similarly. Differences in terms of
information criteria (aic, bic) show that fractional models using different
power indices are also very close to each other. Comparing Tables 5 and 6,
the highest lag in information criteria between both power indices is
registered for the specification where adjusted budget shares are explained
and heteroskedascitity is modelled. In this particular case, the nucleolus
performs better.

Regarding enlargement, two interesting effects are noted. First, the EU
dummy variables show that budget shares decrease with the EU size. Thus, the
larger the number of country members, the lower the average budget share.
Second, power interactions with temporal dummies show that there are
different impacts of power in different subperiods, especially when the
nucleolus is included in the specifications. In general, the period
2004-2006 (EU25) is one where country members achieve higher relative
returns from their political power. When the number of country members
experience a substantial increase, those countries with higher power levels
obtain higher relative gains.

\begin{table}[p]
\caption{Total budget shares ($exp$) and the SSI ($p_{ssi}$) }\centering
{\normalsize {\small \centering
{\ \renewcommand{\arraystretch}{0.80} 
\begin{tabular}{lcccc}
\hline\hline
& \multicolumn{1}{c}{(1)} & \multicolumn{1}{c}{(2)} & \multicolumn{1}{c}{(3)}
& \multicolumn{1}{c}{(4)} \\ 
& \multicolumn{1}{c}{OLS} & \multicolumn{1}{c}{$\text{OLS}_\text{d}$} & 
\multicolumn{1}{c}{GLM} & \multicolumn{1}{c}{FHETPROB} \\ \hline
p\_ssi & 0.545\rlap{$^{**}$} & 0.493\rlap{$^{**}$} & 3.781\rlap{$^{**}$} & 
3.675\rlap{$^{**}$} \\ 
agri & 0.352\rlap{$^{**}$} & 0.337\rlap{$^{**}$} & 1.682\rlap{$^{**}$} & 
1.768\rlap{$^{**}$} \\ 
income & -0.005\rlap{$^{*}$} & -0.004\rlap{$^{+}$} & -0.157\rlap{$^{+}$} & 
-0.263\rlap{$^{**}$} \\ 
p\_ssiEU10 &  & 0.034 & 0.490 & 0.659 \\ 
p\_ssiEU12 &  & 0.001 & 1.305 & 1.401 \\ 
p\_ssiEU15 &  & 0.199\rlap{$^{**}$} & 3.213\rlap{$^{*}$} & 1.787 \\ 
p\_ssiEU25 &  & 0.248\rlap{$^{**}$} & 7.293\rlap{$^{**}$} & 2.902 \\ 
p\_ssiEU27 &  & 0.150\rlap{$^{*}$} & 7.128\rlap{$^{**}$} & 2.096 \\ 
EU10 &  & -0.006 & -0.075 & -0.103 \\ 
EU12 &  & -0.002 & -0.098 & -0.143 \\ 
EU15 &  & -0.014\rlap{$^{*}$} & -0.233 & -0.103 \\ 
EU25 &  & -0.014\rlap{$^{*}$} & -0.482\rlap{$^{*}$} & -0.142 \\ 
EU27 &  & -0.010\rlap{$^{+}$} & -0.450\rlap{$^{*}$} & -0.034 \\ 
tobs32 &  &  &  & -0.569 \\ 
tobs27 &  &  &  & 0.540\rlap{$^{**}$} \\ 
tobs18 &  &  &  & -4.400\rlap{$^{**}$} \\ 
tobs9 &  &  &  & 0.665\rlap{$^{**}$} \\ 
tobs6 &  &  &  & 0.918\rlap{$^{**}$} \\ 
\_cons & 0.006\rlap{$^{*}$} & 0.012\rlap{$^{+}$} & -1.830\rlap{$^{**}$} & 
-1.722\rlap{$^{**}$} \\ \hline
lnsigma2 &  &  &  &  \\ 
tobs32 &  &  &  & 0.370 \\ 
tobs27 &  &  &  & -0.419\rlap{$^{**}$} \\ 
tobs18 &  &  &  & 1.053\rlap{$^{**}$} \\ 
tobs9 &  &  &  & -0.711\rlap{$^{**}$} \\ 
tobs6 &  &  &  & -1.112\rlap{$^{**}$} \\ \hline
$cluster$ & no & no & yes & yes \\ 
$N$ & 575 & 575 & 575 & 575 \\ 
$\text{R}^{2}$\_{adj} & 0.88 & 0.88 &  &  \\ 
chi2 &  &  & 580.92 &  \\ 
aic & -2940.72 & -2945.21 & 200.09 & 269.57 \\ 
bic & -2923.30 & -2884.25 & 261.05 & 356.66 \\ \hline\hline
\multicolumn{5}{l}{{\footnotesize \sym{$^{+}$} $p<0.10$, \sym{$^{*}$}$p<0.05$%
, \sym{$^{**}$} $p<0.01$}} \\ 
&  &  &  & 
\end{tabular}%
} } }
\end{table}

\begin{table}[p]
\caption{Total budget shares ($exp$) and nucleolus ($p_{nucl}$)}\centering
{\normalsize {\small \centering
{\ \renewcommand{\arraystretch}{0.80} 
\begin{tabular}{lcccc}
\hline\hline
& \multicolumn{1}{c}{(1)} & \multicolumn{1}{c}{(2)} & \multicolumn{1}{c}{(3)}
& \multicolumn{1}{c}{(4)} \\ 
& \multicolumn{1}{c}{OLS} & \multicolumn{1}{c}{$\text{OLS}_{d}$} & 
\multicolumn{1}{c}{GLM} & \multicolumn{1}{c}{FHETPROB} \\ \hline
p\_nucl & 0.221\rlap{$^{**}$} & 0.204\rlap{$^{**}$} & 1.369\rlap{$^{*}$} & 
1.345\rlap{$^{*}$} \\ 
agri & 0.504\rlap{$^{**}$} & 0.370\rlap{$^{**}$} & 1.931\rlap{$^{**}$} & 
1.981\rlap{$^{**}$} \\ 
income & -0.010\rlap{$^{**}$} & -0.005\rlap{$^{*}$} & -0.168\rlap{$^{+}$} & 
-0.277\rlap{$^{**}$} \\ 
p\_nuclEU10 &  & 0.037 & 0.471\rlap{$^{+}$} & 0.592\rlap{$^{*}$} \\ 
p\_nuclEU12 &  & 0.199\rlap{$^{**}$} & 3.041\rlap{$^{**}$} & 3.233%
\rlap{$^{**}$} \\ 
p\_nuclEU15 &  & 0.463\rlap{$^{**}$} & 5.574\rlap{$^{**}$} & 3.966%
\rlap{$^{**}$} \\ 
p\_nuclEU25 &  & 0.528\rlap{$^{**}$} & 10.102\rlap{$^{**}$} & 5.008%
\rlap{$^{*}$} \\ 
p\_nuclEU27 &  & 0.419\rlap{$^{**}$} & 9.813\rlap{$^{**}$} & 4.141%
\rlap{$^{*}$} \\ 
EU10 &  & -0.009\rlap{$^{+}$} & -0.108\rlap{$^{*}$} & -0.134\rlap{$^{**}$}
\\ 
EU12 &  & -0.026\rlap{$^{**}$} & -0.323\rlap{$^{**}$} & -0.374\rlap{$^{**}$}
\\ 
EU15 &  & -0.043\rlap{$^{**}$} & -0.507\rlap{$^{**}$} & -0.364\rlap{$^{*}$}
\\ 
EU25 &  & -0.043\rlap{$^{**}$} & -0.770\rlap{$^{**}$} & -0.388\rlap{$^{+}$}
\\ 
EU27 &  & -0.039\rlap{$^{**}$} & -0.731\rlap{$^{**}$} & -0.280 \\ 
tobs32 &  &  &  & -0.014 \\ 
tobs27 &  &  &  & 0.599\rlap{$^{**}$} \\ 
tobs18 &  &  &  & -3.451\rlap{$^{**}$} \\ 
tobs9 &  &  &  & 0.705\rlap{$^{**}$} \\ 
tobs6 &  &  &  & 0.889\rlap{$^{**}$} \\ 
\_cons & 0.022\rlap{$^{**}$} & 0.041\rlap{$^{**}$} & -1.561\rlap{$^{**}$} & 
-1.454\rlap{$^{**}$} \\ \hline
lnsigma2 &  &  &  &  \\ 
tobs32 &  &  &  & 0.083 \\ 
tobs27 &  &  &  & -0.483\rlap{$^{**}$} \\ 
tobs18 &  &  &  & 0.889\rlap{$^{**}$} \\ 
tobs9 &  &  &  & -0.757\rlap{$^{**}$} \\ 
tobs6 &  &  &  & -1.095\rlap{$^{**}$} \\ \hline
$cluster$ & no & no & yes & yes \\ 
$N$ & 575 & 575 & 575 & 575 \\ 
$\text{R}^{2}$\_{adj} & 0.84 & 0.87 &  &  \\ 
chi2 &  &  & 641.14 &  \\ 
aic & -2808.05 & -2912.11 & 200.30 & 269.76 \\ 
bic & -2790.63 & -2851.15 & 261.26 & 356.85 \\ \hline\hline
\multicolumn{5}{l}{{\footnotesize \sym{$^{+}$} $p<0.10$, \sym{$^{*}$}$p<0.05$%
, \sym{$^{**}$} $p<0.01$}} \\ 
&  &  &  & 
\end{tabular}%
} } }
\end{table}

\begin{table}[p]
\caption{Adjusted budget shares ($exp_{adj}$) and the SSI ($p_{ssi}$)}%
\centering
{\normalsize {\small \centering
{\ \renewcommand{\arraystretch}{0.80} 
\begin{tabular}{lcccc}
\hline\hline
& \multicolumn{1}{c}{(1)} & \multicolumn{1}{c}{(2)} & \multicolumn{1}{c}{(3)}
& \multicolumn{1}{c}{(4)} \\ 
& \multicolumn{1}{c}{OLS} & \multicolumn{1}{c}{$\text{OLS}_\text{d}$} & 
\multicolumn{1}{c}{GLM} & \multicolumn{1}{c}{FHETPROB} \\ \hline
p\_ssi & 0.768\rlap{$^{**}$} & 0.733\rlap{$^{**}$} & 5.148\rlap{$^{**}$} & 
5.143\rlap{$^{**}$} \\ 
agri & 0.200\rlap{$^{**}$} & 0.157\rlap{$^{**}$} & 0.704\rlap{$^{*}$} & 0.718%
\rlap{$^{*}$} \\ 
income & -0.002 & 0.001 & -0.103 & -0.223\rlap{$^{*}$} \\ 
p\_ssiEU10 &  & 0.034 & 0.456 & 0.493 \\ 
p\_ssiEU12 &  & 0.080 & 1.876\rlap{$^{*}$} & 2.058\rlap{$^{*}$} \\ 
p\_ssiEU15 &  & 0.326\rlap{$^{**}$} & 4.070\rlap{$^{**}$} & 2.735 \\ 
p\_ssiEU25 &  & 0.316\rlap{$^{**}$} & 7.913\rlap{$^{**}$} & 4.256%
\rlap{$^{+}$} \\ 
p\_ssiEU27 &  & 0.210\rlap{$^{**}$} & 7.571\rlap{$^{**}$} & 3.132 \\ 
EU10 &  & -0.005 & -0.068 & -0.071 \\ 
EU12 &  & -0.007 & -0.141 & -0.192 \\ 
EU15 &  & -0.020\rlap{$^{**}$} & -0.281 & -0.153 \\ 
EU25 &  & -0.012\rlap{$^{+}$} & -0.490\rlap{$^{*}$} & -0.188 \\ 
EU27 &  & -0.008 & -0.443\rlap{$^{*}$} & -0.060 \\ 
tobs32 &  &  &  & -4.990 \\ 
tobs27 &  &  &  & 0.490\rlap{$^{**}$} \\ 
tobs18 &  &  &  & -5.168\rlap{$^{**}$} \\ 
tobs9 &  &  &  & 0.414\rlap{$^{**}$} \\ 
tobs6 &  &  &  & 0.931\rlap{$^{**}$} \\ 
\_cons & -0.002 & 0.000 & -1.926\rlap{$^{**}$} & -1.808\rlap{$^{**}$} \\ 
\hline
lnsigma2 &  &  &  &  \\ 
tobs32 &  &  &  & 1.454 \\ 
tobs27 &  &  &  & -0.376\rlap{$^{**}$} \\ 
tobs18 &  &  &  & 1.163\rlap{$^{**}$} \\ 
tobs9 &  &  &  & -0.491\rlap{$^{**}$} \\ 
tobs6 &  &  &  & -1.091\rlap{$^{**}$} \\ \hline
$cluster$ & no & no & yes & yes \\ 
$N$ & 575 & 575 & 575 & 575 \\ 
$\text{R}^{2}$\_{adj} & 0.87 & 0.88 &  &  \\ 
chi2 &  &  & 1121.58 &  \\ 
aic & -2926.94 & -2964.45 & 199.72 & 271.41 \\ 
bic & -2909.53 & -2903.49 & 260.68 & 362.85 \\ \hline\hline
\multicolumn{5}{l}{{\footnotesize \sym{$^{+}$} $p<0.10$, \sym{$^{*}$}$p<0.05$%
, \sym{$^{**}$} $p<0.01$}} \\ 
&  &  &  & 
\end{tabular}%
} } }
\end{table}

\begin{table}[p]
\caption{Adjusted budget shares ($exp_{adj}$) and nucleolus ($p_{nucl}$)}%
\centering
{\normalsize {\small \centering
{\ \renewcommand{\arraystretch}{0.80} 
\begin{tabular}{lcccc}
\hline\hline
& \multicolumn{1}{c}{(1)} & \multicolumn{1}{c}{(2)} & \multicolumn{1}{c}{(3)}
& \multicolumn{1}{c}{(4)} \\ 
& \multicolumn{1}{c}{OLS} & \multicolumn{1}{c}{$\text{OLS}_\text{d}$} & 
\multicolumn{1}{c}{GLM} & \multicolumn{1}{c}{FHETPROB} \\ \hline
p\_nucl & 0.324\rlap{$^{**}$} & 0.318\rlap{$^{**}$} & 2.015\rlap{$^{**}$} & 
2.018\rlap{$^{**}$} \\ 
agri & 0.404\rlap{$^{**}$} & 0.198\rlap{$^{**}$} & 0.965\rlap{$^{*}$} & 0.974%
\rlap{$^{*}$} \\ 
income & -0.009\rlap{$^{**}$} & -0.001 & -0.114 & -0.251\rlap{$^{**}$} \\ 
p\_nuclEU10 &  & 0.039 & 0.477\rlap{$^{+}$} & 0.602\rlap{$^{*}$} \\ 
p\_nuclEU12 &  & 0.362\rlap{$^{**}$} & 4.206\rlap{$^{**}$} & 4.463%
\rlap{$^{**}$} \\ 
p\_nuclEU15 &  & 0.719\rlap{$^{**}$} & 7.274\rlap{$^{**}$} & 5.688%
\rlap{$^{**}$} \\ 
p\_nuclEU25 &  & 0.726\rlap{$^{**}$} & 11.574\rlap{$^{**}$} & 7.024%
\rlap{$^{**}$} \\ 
p\_nuclEU27 &  & 0.604\rlap{$^{**}$} & 11.070\rlap{$^{**}$} & 5.840%
\rlap{$^{**}$} \\ 
EU10 &  & -0.010\rlap{$^{+}$} & -0.114\rlap{$^{*}$} & -0.141\rlap{$^{**}$}
\\ 
EU12 &  & -0.041\rlap{$^{**}$} & -0.439\rlap{$^{**}$} & -0.493\rlap{$^{**}$}
\\ 
EU15 &  & -0.062\rlap{$^{**}$} & -0.646\rlap{$^{**}$} & -0.499\rlap{$^{*}$}
\\ 
EU25 &  & -0.055\rlap{$^{**}$} & -0.866\rlap{$^{**}$} & -0.505\rlap{$^{*}$}
\\ 
EU27 &  & -0.050\rlap{$^{**}$} & -0.810\rlap{$^{**}$} & -0.379\rlap{$^{+}$}
\\ 
tobs32 &  &  &  & -0.217 \\ 
tobs27 &  &  &  & 0.589\rlap{$^{**}$} \\ 
tobs18 &  &  &  & -3.597\rlap{$^{**}$} \\ 
tobs9 &  &  &  & 0.464\rlap{$^{**}$} \\ 
tobs6 &  &  &  & 0.890\rlap{$^{**}$} \\ 
\_cons & 0.020\rlap{$^{**}$} & 0.043\rlap{$^{**}$} & -1.574\rlap{$^{**}$} & 
-1.439\rlap{$^{**}$} \\ \hline
lnsigma2 &  &  &  &  \\ 
tobs32 &  &  &  & 0.193 \\ 
tobs27 &  &  &  & -0.487\rlap{$^{**}$} \\ 
tobs18 &  &  &  & 0.909\rlap{$^{**}$} \\ 
tobs9 &  &  &  & -0.547\rlap{$^{**}$} \\ 
tobs6 &  &  &  & -1.083\rlap{$^{**}$} \\ \hline
$cluster$ & no & no & yes & yes \\ 
$N$ & 575 & 575 & 575 & 575 \\ 
$\text{R}^{2}$\_{adj} & 0.81 & 0.87 &  &  \\ 
chi2 &  &  & 1283.98 &  \\ 
aic & -2704.59 & -2906.01 & 200.04 & 269.71 \\ 
bic & -2687.18 & -2845.05 & 261.00 & 356.80 \\ \hline\hline
\multicolumn{5}{l}{{\footnotesize \sym{$^{+}$} $p<0.10$, \sym{$^{*}$}$p<0.05$%
, \sym{$^{**}$} $p<0.01$}} \\ 
&  &  &  & 
\end{tabular}%
} } }
\end{table}

As mentioned earlier, Table 7 provides marginal effects, allowing the
comparision of all models and specifications in terms of impact of power and
needs. OLS techniques tend to overestimate the value of the coefficients,
while fractional models smooth the estimates, since they are more
appropriate when modelling shares. Another interesting idea emerges where
power indices have a higher impact on adjusted budget share specifications.
According to our expectations, once the budget shares are adjusted to take
the UK rebate and other compensation payments into consideration, the
explanatory weight of power increases. Nevertheless, the relative weight of
political power is lower when considering the nucleolus, while needs'
factors become more important. There is no doubt that political power
matters, but not as much as the models with the SSI as a power index have
shown. Thus, the specifications based on the nucleolus show a more balanced
situation between power and needs. The result reinforces the idea that the
impact of political power on budget share is not as important as \cite%
{kauppiwidgren04} predicted.\footnote{%
In this respect, the original estimates of the SSI by \cite{kauppiwidgren04}
for the period 1976-2001 (considering annual data) ranged between 0.545 and
0.645 and between 0.783 and 0.858 for the total budget shares and adjusted
budget shares respectively. The impact of agriculture was estimated in a
range between 0.387 and 0.405 and between 0.252 and 0.236 for the total
budget shares and adjusted budget shares respectively. Finally, estimated
income coefficient registered values between -0.025 and 0.008 and between
-0.022 and 0.003 for the total budget shares and adjusted budget shares
respectively.}

\bigskip

\FloatBarrier
\begin{table}[h]
\caption{Power versus needs: marginal effects}\centering
{\normalsize {\small \centering
{\ \renewcommand{\arraystretch}{0.80} 
\begin{tabular}{lcccc}
\hline\hline
& \multicolumn{1}{c}{OLS} & \multicolumn{1}{c}{$\text{OLS}_\text{d}$} & 
\multicolumn{1}{c}{GLM} & \multicolumn{1}{c}{FHETPROB} \\ \hline
\multicolumn{1}{l}{Total budget share/SSI} &  &  &  &  \\ \hline
p\_ssi & 0.545\rlap{$^{**}$} & 0.493\rlap{$^{**}$} & 0.345\rlap{$^{**}$} & 
0.426\rlap{$^{**}$} \\ 
agri & 0.352\rlap{$^{**}$} & 0.337\rlap{$^{**}$} & 0.153\rlap{$^{**}$} & 
0.205\rlap{$^{**}$} \\ 
income & -0.005\rlap{$^{*}$} & -0.004\rlap{$^{+}$} & -0.014\rlap{$^{+}$} & 
-0.031\rlap{$^{**}$} \\ \hline
\multicolumn{5}{l}{Total budget share/nucleolus} \\ \hline
p\_nucl & 0.221\rlap{$^{**}$} & 0.204\rlap{$^{**}$} & 0.125\rlap{$^{*}$} & 
0.161\rlap{$^{*}$} \\ 
agri & 0.504\rlap{$^{**}$} & 0.370\rlap{$^{**}$} & 0.176\rlap{$^{**}$} & 
0.237\rlap{$^{**}$} \\ 
income & -0.010\rlap{$^{**}$} & -0.005\rlap{$^{*}$} & -0.014\rlap{$^{+}$} & 
-0.033\rlap{$^{**}$} \\ \hline
\multicolumn{5}{l}{Adjusted budget share/SSI} \\ \hline
p\_ssi & 0.768\rlap{$^{**}$} & 0.733\rlap{$^{**}$} & 0.466\rlap{$^{**}$} & 
0.563\rlap{$^{**}$} \\ 
agri & 0.200\rlap{$^{**}$} & 0.157\rlap{$^{**}$} & 0.064\rlap{$^{*}$} & 0.079%
\rlap{$^{*}$} \\ 
income & -0.002 & 0.001 & -0.009 & -0.024\rlap{$^{*}$} \\ \hline
\multicolumn{5}{l}{Adjusted budget share/nucleolus} \\ \hline
p\_nucl & 0.324\rlap{$^{**}$} & 0.318\rlap{$^{**}$} & 0.183\rlap{$^{**}$} & 
0.235\rlap{$^{**}$} \\ 
agri & 0.404\rlap{$^{**}$} & 0.198\rlap{$^{**}$} & 0.088\rlap{$^{*}$} & 0.114%
\rlap{$^{*}$} \\ 
income & -0.009\rlap{$^{**}$} & -0.001 & -0.010 & -0.029\rlap{$^{**}$} \\ 
\hline\hline
\multicolumn{5}{l}{{\footnotesize \sym{$^{+}$} $p<0.10$, \sym{$^{*}$}$p<0.05$%
, \sym{$^{**}$} $p<0.01$}} \\ 
&  &  &  & 
\end{tabular}%
} } }
\end{table}
\FloatBarrier

\bigskip \bigskip \bigskip \bigskip \bigskip

\section{Discussion and future research}

The main contribution of this paper is to highlight the role of political
power on the EU budget decisions. Various key drivers of budget shares
allocated to each EU member country have been identified. Both power and
needs are significant factors in explaining expenditure budget allocation
among EU member states. Some previous empirical analysis %
\citep{kauppiwidgren04, kauppiwidgren07} show strong prevalence of political
power motives. Their results indicate that a large percentage of budget
expenditures can be attributed to selfish power politics, leaving a small
contribution to the so-called benevolent EU need-based budget policies.

We have carried out an empirical analysis to revisit the findings of \cite%
{kauppiwidgren04}. To this end, we have updated their data set (originally
from 1976 to 2001, the range has been extended to 2012). Additionally we
have compared alternative political power measures and have applied more
sophisticated and rigorous econometric techniques. We have argued that the
nucleolus \citep{schmeidler69} is a good alternative to the SSI when
explaining the budget share of EU member states, from both a theoretical and
an empirical perspective.

Our findings show that under simple econometric specifications, both power
indices behave in a similar way, although the SSI is slightly superior in
terms of explanatory power. However, when using more sophisticated and
adequate econometric techniques, the nucleolus seems to perform better than
the SSI.\footnote{%
In general, the voting power need not be proportional to the voting weights %
\citep{felsenthal98}. However, following a na\"{\i}ve approach, we have also
performed estimations using the voting weights instead of the power indices.
The results imply that the specifications based on the power indices perform
better. More details are available from the authors upon request.} In
particular, the nucleolus performs better when considering adjusted budget
shares (by compensations, such as the UK rebate), and when we adjust for the
unbalancedness of the panel data. Moreover, the higher the number of
countries competing for EU budget, the higher the impact of political power
on budget shares.

Additionally, we find that the relative weight of political power based on
the nucleolus when explaining budget shares is lower than predictions of
other models. A partial explanation may be the fact that the nucleolus can
assign zero power to non dummy players, which is not the case for the
Shapley-Shubik and the Banzhaf indices. Indeed, this occurs for all voting
systems until 1994 (see Appendix 1). Needs also matter, and countries with
lower relative income levels and a broader irrigation sector are recipients
of a significant amount of EU resources. These findings are consistent with
the idea that the EU budget is allocated to develop key policies such as the
common agricultural policy and the structural funds. Although political
power has an impact on the EU budgetary decisions, this impact seems to be
more moderate than estimated in previous literature. Definitively, the
solidarity principle emerges as a significant key driver of the EU budgetary
allocation.

Finally, we would like to pursue this line of research further through a
more focused analysis of specific sections of the EU budget. When modelling
bargaining schemes, interactions among different sections of the EU budget
will be considered. Additionally, further empirical analysis will aim to
detect factors that increase the probability of receiving higher budget
shares for specific policies.

\section*{Acknowledgements}

We are grateful to professor Heikki Kauppi, who shared the original data set
(1976-2001) with us. We would also like to thank Michel Le Breton, Mar\'{\i}%
a Montero, Fran\c{c}ois Salani\'{e} and Roberto Mart\'{\i}nez Espi\~{n}eira
for their input during insightful discussions at the early stages of our
work, and the Editor and two anonymous referees for their comments and
suggestions.

\begin{landscape}
\section*{\normalsize{Appendix 1: The SSI and the nucleolus in the Council of Ministers
1958-2002}}
\small{
\begin{tabular}{|p{3.5cm}|p{1cm}|p{1cm}|p{1cm}|p{1cm}|p{1cm}|p{1cm}||p{1cm}||p{1cm}||p{1cm}||p{1cm}|}
\hline\hline
\textbf{Country} & \multicolumn{2}{|c|}{$\mathbf{1958-1972}$} & 
\multicolumn{2}{|c|}{$\mathbf{1973-1980}$} & \multicolumn{2}{|c|}{$\mathbf{1981-1985}$} & \multicolumn{2}{|c|}{$\mathbf{1986-1994}$} & 
\multicolumn{2}{|c|}{$\mathbf{1995-2002}$} \\ \hline
& \textbf{SSI} & \textbf{Nucl} & \textbf{SSI} & \textbf{Nucl} & \textbf{SSI}
& \textbf{Nucl} & \textbf{SSI} & \textbf{Nucl} & \textbf{SSI} & \textbf{Nucl}
\\ \hline
\textbf{France} & $0.233$ & \multicolumn{1}{|r|}{$\mathbf{0.250}$} & $0.179$
& $\mathbf{0.250}$ & $0.174$ & $\mathbf{0.250}$ & $0.134$ & $\mathbf{0.138}$
& $0.117$ & $\mathbf{0.115}$ \\ 
\textbf{Germany} & $0.233$ & \multicolumn{1}{|r|}{$\mathbf{0.250}$} & $0.179$
& $\mathbf{0.250}$ & $0.174$ & $\mathbf{0.250}$ & $0.134$ & $\mathbf{0.138}$
& $0.117$ & $\mathbf{0.115}$ \\ 
\textbf{Italy} & $0.233$ & \multicolumn{1}{|r|}{$\mathbf{0.250}$} & $0.179$
& $\mathbf{0.250}$ & $0.174$ & $\mathbf{0.250}$ & $0.134$ & $\mathbf{0.138}$
& $0.117$ & $\mathbf{0.115}$ \\
\textbf{United Kingdom} & $-$ & $\mathbf{-}$ & $0.179$ & $\mathbf{0.250}$ & $0.174$ & $\mathbf{0.250}$ & $0.134$ & $\mathbf{0.138}$ & $0.117$ & $\mathbf{0.115}$ \\  
\textbf{Belgium} & $0.150$ & \multicolumn{1}{|r|}{$\mathbf{0.125}$} & $0.081$
& $\mathbf{0}$ & $0.071$ & $\mathbf{0}$ & $0.064$ & \multicolumn{1}{|r|}{$\mathbf{0.069}$} & $0.056$ & $\mathbf{0.057}$ \\ 
\textbf{Netherlands} & $0.150$ & \multicolumn{1}{|r|}{$\mathbf{0.125}$} & $0.081$ & $\mathbf{0}$ & $0.071$ & $\mathbf{0}$ & $0.064$ & 
\multicolumn{1}{|r|}{$\mathbf{0.069}$} & $0.056$ & $\mathbf{0.057}$
\\ 
\textbf{Luxembourg} & $0$ & $\mathbf{0}$ & $0.001$ & $\mathbf{0}$ & $0.030$ & $\mathbf{0}$ & $0.012$ & $\mathbf{0}$ & $0.021$ & 
\multicolumn{1}{|r|}{$\mathbf{0.023}$} \\ 
\textbf{Denmark} & $-$ & $\mathbf{-}$ & $0.057$ & $\mathbf{0}$ & $0.030$ & $\mathbf{0}$ & $0.043$ & $\mathbf{0.034}$ & $0.035$ & $\mathbf{0.034}$ \\ 
\textbf{Ireland} & $-$ & $\mathbf{-}$ & $0.057$ & $\mathbf{0}$ & $0.030$ & $\mathbf{0}$ & $0.043$ & $\mathbf{0.034}$ & $0.035$ & $\mathbf{0.034}$ \\ 
\textbf{Greece} & $-$ & $\mathbf{-}$ & $-$ & $\mathbf{-}$ & $0.071$ & $\mathbf{0}$ & $0.064$ & \multicolumn{1}{|r|}{$\mathbf{0.069}$} & $0.056$ & $\mathbf{0.057}$ \\ 
\textbf{Spain} & $-$ & $\mathbf{-}$ & $-$ & $\mathbf{-}$ & $-$ & $\mathbf{-}$
& $0.111$ & $\mathbf{0.103}$ & $0.095$ & $\mathbf{0.092}$ \\ 
\textbf{Portugal} & $-$ & $\mathbf{-}$ & $-$ & $\mathbf{-}$ & $-$ & $\mathbf{-}$ & $0.064$ & $\mathbf{0.069}$ & $0.056$ & $\mathbf{0.057}$ \\ 
\textbf{Austria} & $-$ & $\mathbf{-}$ & $-$ & $\mathbf{-}$ & $-$ & $\mathbf{-}$ & $-$ & $\mathbf{-}$ & $0.045$ & $\mathbf{0.046}$ \\ 
\textbf{Sweden} & $-$ & $\mathbf{-}$ & $-$ & $\mathbf{-}$ & $-$ & $\mathbf{-}
$ & $-$ & $\mathbf{-}$ & $0.045$ & $\mathbf{0.046}$ \\ 
\textbf{Finland} & $-$ & $\mathbf{-}$ & $-$ & $\mathbf{-}$ & $-$ & $\mathbf{-}$ & $-$ & $\mathbf{-}$ & $0.035$ & $\mathbf{0.034}$ \\ \hline
\multicolumn{11}{l}{ \textit Source: Adapted from Le Breton et al. (2012).}\\
\end{tabular}}

\begin{table}[H]\centering\section*{\normalsize{Appendix 1 (cont.): The SSI and the nucleolus
in the Council of Ministers under the Treaty of Nice, 2003-2012}}\centering
\tiny{
\begin{tabular}{|p{4cm}|p{1.9cm}|p{1.9cm}|p{1.9cm}|p{1.9cm}|p{1.9cm}|p{1.9cm}|}
\hline\hline
\textbf{Country} & \multicolumn{2}{|c|}{$\mathbf{2003}$} & 
\multicolumn{2}{|c|}{$\mathbf{2004-2006}$} & \multicolumn{2}{|c|}{$\mathbf{2007-2012}$} \\ \hline
& \textbf{SSI} & \textbf{Nucl} & \textbf{SSI} & \textbf{Nucl} & \textbf{SSI}
& \textbf{Nucl} \\ \hline
\textbf{France} & $0.129$ & $\mathbf{0.122}$ & $0.094$ & $\mathbf{0.090}$ & $0.087$ & $\mathbf{0.084}$ \\ 
\textbf{Germany} & $0.137$ & $\mathbf{0.122}$ & $0.095$ & $\mathbf{0.090}$ & 
$0.088$ & $\mathbf{0.084}$ \\ 
\textbf{Italy} & $0.128$ & $\mathbf{0.122}$ & $0.094$ & $\mathbf{0.090}$ & $0.087$ & $\mathbf{0.084}$ \\ 
\textbf{United Kingdom} & $0.129$ & $\mathbf{0.122}$ & $0.094$ & $\mathbf{0.090}$ & $0.087$ & $\mathbf{0.084}$ \\ 
\textbf{Belgium} & $0.046$ & $\mathbf{0.051}$ & $0.036$ & $\mathbf{0.037}$ & 
$0.034$ & $\mathbf{0.035}$ \\ 
\textbf{Netherlands} & $0.052$ & $\mathbf{0.051}$ & $0.039$ & $\mathbf{0.040}
$ & $0.037$ & $\mathbf{0.038}$ \\ 
\textbf{Denmark} & $0.028$ & $\mathbf{0.031}$ & $0.021$ & $\mathbf{0.022}$ & 
$0.019$ & $\mathbf{0.020}$ \\ 
\textbf{Ireland} & $0.028$ & $\mathbf{0.031}$ & $0.021$ & $\mathbf{0.022}$ & 
$0.019$ & $\mathbf{0.020}$ \\ 
\textbf{Luxembourg} & $0.016$ & $\mathbf{0.020}$ & $0.012$ & $\mathbf{0.012}$
& $0.011$ & $\mathbf{0.012}$ \\ 
\textbf{Greece} & $0.046$ & $\mathbf{0.051}$ & $0.036$ & $\mathbf{0.037}$ & $0.034$ & $\mathbf{0.035}$ \\ 
\textbf{Portugal} & $0.046$ & $\mathbf{0.051}$ & $0.036$ & $\mathbf{0.037}$
& $0.034$ & $\mathbf{0.035}$ \\ 
\textbf{Spain} & $0.110$ & $\mathbf{0.112}$ & $0.087$ & $\mathbf{0.084}$ & $0.080$ & $\mathbf{0.078}$ \\ 
\textbf{Austria} & $0.038$ & $\mathbf{0.041}$ & $0.030$ & $\mathbf{0.031}$ & 
$0.028$ & $\mathbf{0.029}$ \\ 
\textbf{Finland} & $0.028$ & $\mathbf{0.031}$ & $0.021$ & $\mathbf{0.022}$ & 
$0.019$ & $\mathbf{0.020}$ \\ 
\textbf{Sweden} & $0.038$ & $\mathbf{0.041}$ & $0.030$ & $\mathbf{0.031}$ & $0.028$ & $\mathbf{0.029}$ \\ 
\textbf{Cyprus} & $-$ & $-$ & $0.012$ & $\mathbf{0.012}$ & $0.011$ & $\mathbf{0.012}$ \\ 
\textbf{Slovakia} & $-$ & $-$ & $0.021$ & $\mathbf{0.022}$ & $0.019$ & $\mathbf{0.020}$ \\ 
\textbf{Slovenia} & $-$ & $-$ & $0.012$ & $\mathbf{0.012}$ & $0.011$ & $\mathbf{0.012}$ \\ 
\textbf{Estonia} & $-$ & $-$ & $0.012$ & $\mathbf{0.012}$ & $0.011$ & $\mathbf{0.012}$ \\ 
\textbf{Hungary} & $-$ & $-$ & $0.036$ & $\mathbf{0.037}$ & $0.034$ & $\mathbf{0.035}$ \\ 
\textbf{Latvia} & $-$ & $-$ & $0.012$ & $\mathbf{0.012}$ & $0.011$ & $\mathbf{0.012}$ \\ 
\textbf{Lithuania} & $-$ & $-$ & $0.021$ & $\mathbf{0.022}$ & $0.019$ & $\mathbf{0.020}$ \\ 
\textbf{Malta} & $-$ & $-$ & $0.009$ & $\mathbf{0.009}$ & $0.008$ & $\mathbf{0.009}$ \\ 
\textbf{Poland} & $-$ & $-$ & $0.087$ & $\mathbf{0.084}$ & $0.080$ & $\mathbf{0.078}$ \\ 
\textbf{Czech Republic} & $-$ & $-$ & $0.036$ & $\mathbf{0.037}$ & $0.034$ & 
$\mathbf{0.035}$ \\ 
\textbf{Bulgaria} & $-$ & $-$ & $-$ & $-$ & $0.028$ & $\mathbf{0.029}$ \\ 
\textbf{Romania} & $-$ & $-$ & $-$ & $-$ & $0.040$ & $\mathbf{0.041}$ \\ 
\hline
\multicolumn{7}{l}{ \textit Source: Own elaboration.}\\
\end{tabular}}
\end{table}
\end{landscape}%

\section*{Appendix 2}

\subsection{Technical preliminaries}

In this section, we introduce basic notions commonly used to model voting
situations, and then briefly discuss the nucleolus and the Shapley-Shubik
index.

We consider a set $N=\left\{ 1,...,n\right\} $ of $n$ players or voters,
which is often referred to as an assembly. The power set $2^{N}$ contains
all the subsets of $N$. A non-empty subset $S\subseteq N$ is called a
coalition. The coalition $N$ is said to be the grand coalition.

A cooperative game with transferable utility in characteristic function
form, is a pair\ $\left( N,v\right) $ with $N$ the set of players and: 
\begin{equation*}
v:2^{N}\longrightarrow \mathbb{R\;}\mathbf{:\;}S\longmapsto v(S),
\end{equation*}%
which is a map that satisfies $v(\emptyset )=0$. The map $v$ is called the
characteristic function. The value $v(S)$ is said to be the value or the
worth of coalition $S$. For simplicity, we refer to these games as
\textquotedblleft games in TU form.\textquotedblright

The game $\left( N,v\right) $ is said to be simple if:

$\cdot $ the value of a coalition is either $0$ or $1$: $v(S)\in \left\{
0,1\right\} $ for all $S\subseteq N$,

$\cdot $ the value of grand coalition is $1$: $v(N)=1$.

A coalition with a value equal to $1$ is said to be winning, and a coalition
with a value equal to $0$ is said to be losing. A winning coalition $S$ is
minimal if it does not contain any other winning coalition: $v(S)=1$ and $%
v(T)=0$ for all $T\subset S$. Furthermore, the set of winning coalitions is
denoted by $\mathcal{W}$ and the set of minimal winning coalitions\ is
denoted by $\mathcal{W}^{m}$. The simple game $\left( N,v\right) $ is fully
determined through the pair $\left( N,\mathcal{W}\right) $.

Furthermore, the simple game is said to be monotonic if supersets of winning
coalitions are winning, i.e., if $S\in \mathcal{W}$ and $T\supset S$, then $%
T\in \mathcal{W}$. A monotonic simple game is also called a simple voting
game.\footnote{%
Several authors also use the term simple game for simple voting games, i.e.,
where monotonicity is assumed.}

Voting situations are often described by weighted majority games, for
example the one in the EU Council of Ministers. The game $\left( N,v\right) $%
\ is said to be a weighted majority game if there is an $n$-tuple $w=(\omega
_{1},...,\omega _{n})$ of non-negative weights with $\omega _{1}+\omega
_{2}+...+\omega _{n}=1$ and a non-negative quota $q$ such that $v(S)=1$, if
and only if, the total weight of the players in $S$ exceeds the quota $q$,
i.e., 
\begin{equation*}
v(S)=1\text{, if and only if, }\sum_{i\in S}\omega _{i}\geq q.
\end{equation*}%
The pair $\left[ q;\omega \right] $ is called a representation of the game $%
\left( N,v\right) $. Typical examples of weighted majority games are:

$\cdot $ the majority game: $w=(\underset{n}{\underbrace{1,1,...,1}})$ and $%
q=(n+1)/2,$

$\cdot $ the unanimity game: $w=(1,1,...,1)$ and $q=n$,

$\cdot $ the dictator game: $w=(1,0,0,...,0)$ and $q=1$ (player $1$ is the
dictator).

A measure of power is a map $\xi $ from the set of simple voting games $%
\left( N,v\right) $ to the set of $n$-tuples of real numbers. The value $\xi
_{i}=\xi _{i}\left( N,v\right) $ is the power of player $i$ in the game $%
\left( N,v\right) ,$ and it satisfies $0\leq \xi _{i}\leq 1$.

\subsection{Shapley-Shubik Index}

One of the most famous power measures used in the literature is the
Shapley-Shubik index.\footnote{%
For the definitions and properties, see \cite{felsenthal98}.} Several
approaches are used in the literature to present and interpret the
Shapley-Shubik index. \cite{shapley54} apply the following scheme to
introduce their index. The players vote in order and as a majority is
reached, the bill is passed. The critical\footnote{%
Player $i$ in coalition $S$ is said to be \textsl{critical} in $S$ if
without player $i$ the coalition\ left behind is loosing, i.e.,\ 
\begin{equation*}
i~\mathrm{is~critical~in}~S\qquad \mathrm{if}\qquad i\in S\in \mathcal{W}~%
\mathrm{and}~S\setminus \{i\}\notin \mathcal{W}.
\end{equation*}%
If $i$ is not critical in any $S\in \mathcal{W}$, then $i$ is a dummy.}
voter is assumed to be given credit for having passed the bill. The index is
then determined through the assumption of a random voting order.

\begin{definition}
\label{SS}Let $(N,v)$ be a simple voting game. The Shapley-Shubik index
(SSI) of player $i$ is defined by:\begin{equation}
\phi _{i}=\phi _{i}(N,v)=\dsum\limits_{S:\text{ }i\mathcal{\ }\text{is
critical in }S\text{ }}\frac{\left( \left\vert S\right\vert -1\right)
!\left( n-\left\vert S\right\vert \right) !}{n!}\text{ for all }i\in N\text{.}  \label{SSI}
\end{equation}
\end{definition}

The advantage of this approach is that it is simple and non-technical.
However, the authors emphasize that this scheme, \textquotedblleft is just a
convenient conceptual device.\textquotedblright\ The main shortcoming of
this scheme is that this voting model cannot be considered realistic: there
is no reason why the pivot voter should get all the credit, or why the order
of the grand coalition formation should matter.\footnote{%
For more detailed discussion, see \cite{felsenthal98}.} With respect to the
computation the Shapley-Shubik indices of our EU instances, we remark that
looping over all $2^{n}$ coalitions and determining the critical voters was
quick enough for our purpose, i.e., more advanced methods involving
generating functions were not needed.

\subsection{The Nucleolus}

The nucleolus is a solution concept for cooperative games, which was first
formulated by \cite{schmeidler69}. In order to introduce it let us consider
a characteristic function game $\left( N,v\right) $. For convenience, for
some vector $x$ we define: 
\begin{equation*}
x(S)\equiv \sum_{i\in S}x_{i}\text{ for any }S\subseteq N.
\end{equation*}

A payoff vector $x=(x_{1},...,x_{n})$ with $x_{i}\geq v(i)$ and $x(N)=v(N)$
is called an imputation. We denote by $X(N,v)$ the set of all imputations of
the game $(N,v)$.

Let $x$ be an imputation, then for any coalition $S$ the excess of $S$ is
defined as:%
\begin{equation*}
e(S,x)=v(S)-x(S).
\end{equation*}

One might interpret this number as a measure of \textquotedblleft
dissatisfaction\textquotedblright\ for coalition $S$ at imputation $x$. For
any imputation $x$ let $S_{1},...,S_{2^{n}-1}$ be an ordering of the
coalitions for which $e\left( S_{l},x\right) \geq e\left( S_{l+1},x\right) $
for $l=1,...,2^{n}-2$. Let $E(x)$ be the vector of excess defined as $%
E_{l}(x)=e\left( S_{l},x\right) $ for all $l=1,...,2^{n}-1$. We say that $%
E(x)$ is lexicographically less than $E(y)$ if: 
\begin{equation*}
E_{l}(x)<E_{l}(y)\text{ for the smallest }l\text{ for which }E_{l}(x)\neq
E_{l}(y).
\end{equation*}%
We denote this relation by $E(x)\prec _{lex\min }E(y)$.

\begin{definition}
The \textit{nucleolus} is the set of imputations $x$ for which the vector $E(x)$ is lexicographically minimal: 
\begin{equation*}
\nu =\nu (N,v)=\left\{ x\in X(N,v):\text{ }\nexists y\in X(N,v):E(y)\prec
_{lex\min }E(x)\right\} .
\end{equation*}
\end{definition}%

The following recursive procedure is used to characterize the nucleolus. By
definition, $E_{1}\left( x\right) $ is the largest excess of any coalition
relative to $x$. At the first step of the procedure we find the set $X_{1}$
of all imputations $x$ that minimizes $E_{1}\left( x\right) $:%
\begin{equation*}
\begin{array}{l}
\min \epsilon \\ 
\text{s.t. }e(S,x)\leq \epsilon \text{ for all }S,\text{ }\emptyset
\nsubseteqq S\nsubseteqq N \\ 
\text{and }x(N)=v(N).%
\end{array}%
.
\end{equation*}

The set $X_{1}$ is called the least core of $c$. If it is not a unique
point, we find the set $X_{2}$ of all $x$ in $X_{1}$ that minimizes $%
E_{2}\left( x\right) $, the second largest excess and so on. This process
eventually leads to an $X_{k}$ consisting of a single imputation, called the
nucleolus \citep{schmeidler69, maschler79}. The nucleolus recursively
minimizes the \textquotedblright dissatisfaction\textquotedblright\ of the
worst treated coalitions.\footnote{%
Notwithstanding the general recursive definition of the nucleolus, the
computation of $X_{1}$ was sufficient in all EU instances, i.e., a single
linear program has to be solved. The uniqueness of the solution was verified
using the complementary slackness condition.}

It has been proved that the nucleolus of a game in coalitional form exists
and it is unique. Moreover, if the core is not empty, the nucleolus is in
the core \citep{ maschler13}.

\end{document}